# Bias and Angular Dependence of Spin-Transfer Torque in Magnetic Tunnel Junctions

C. Wang[1], Y.-T. Cui[1], J. Z. Sun[2], J. A. Katine[3], R. A. Buhrman[1], and D. C. Ralph[1]

*[1]Cornell University, Ithaca, New York 14853, USA*

*[2]IBM T. J. Watson Research Center, Yorktown Heights, New York 10598, USA*

*[3]Hitachi Global Storage Technologies, San Jose Res. Ctr., San Jose, CA 95135 USA*

We use spin-transfer-driven ferromagnetic resonance (ST-FMR) to measure the spin-transfer torque vector $\tau$ in MgO-based magnetic tunnel junctions as a function of the offset angle between the magnetic moments of the electrodes and as a function of bias, $V$. We explain the conflicting conclusions of two previous experiments by accounting for additional terms that contribute to the ST-FMR signal at large $|V|$. Including the additional terms gives us improved precision in the determination of $\tau(V)$, allowing us to distinguish among competing predictions. We determine that the in-plane component of $d\tau/dV$ has a weak but non-zero dependence on bias, varying by 30-35% over the bias range where the measurements are accurate, and that the perpendicular component can be large enough to be technologically significant. We also make comparisons to other experimental techniques that have been used to try to measure $\tau(V)$.

# INTRODUCTION

MgO-based magnetic tunnel junctions (MTJs) are under intensive investigation for use in memory technologies because of their large magnetoresistance[1,2,3,4] and because their magnetic orientations can be controlled using spin-transfer torques.[5] Determining the magnitude and direction of the spin-torque vector, $\tau$, is fundamental to understanding spin-dependent transport and also for making practical devices. Quantitative measurements of $\tau$ can be made using spin-transfer-driven ferromagnetic resonance (ST-FMR).[6,7,8,9,10] However, two initial experiments using ST-FMR in MgO-based MTJs[11,12] have produced very different results for the dependence of $\tau$ on bias voltage, $V$, and theoretical predictions differ significantly as well.[13,14,15] Our Cornell and IBM co-authors, using MTJs with resistance-area products $RA = 12$ $\Omega$-$\mu m^2$ and with offset angles between the electrode magnetizations, $\theta$, between 50° and 90°, measured that the component of the "torkance" $d\tau/dV$ in the plane defined by the electrode magnetizations (*i.e.*, $d\tau_{\parallel}/dV$) varied by less than 16% for $|V| < 0.3$ V (stated in ref. [11] as $< \pm 8\%$ variation). Kubota et al., using MTJs with $RA = 2$ $\Omega$-$\mu m^2$ and $\theta = 137°$, reported that $d\tau_{\parallel}/dV$ was approximately constant for $|V| < 0.1$ V but had a very asymmetric bias dependence for 0.1 V $< |V| < 0.3$ V, increasing by a factor of 3 for one sign of $V$ and decreasing to change sign for the other (see supplemental Fig. S3d in ref. [12]; Fig. 4 in the main text plots $\tau_{\parallel}$, not $d\tau_{\parallel}/dV$). This very asymmetric bias dependence has been interpreted[12] as support for the predictions of tight binding calculations[13] and can be fit to a scattering formulation.[14] However, *ab initio* calculations predict a weaker bias dependence for $d\tau_{\parallel}/dV$.[15] To resolve these discrepancies, we have performed ST-FMR measurements over a broad range of offset angles and bias for two sets of MTJs with different $RA$ values and have reanalyzed the contributions to the ST-FMR signal. We find that more terms contribute to the high-bias ST-FMR signal than were previously accounted for, modifying the signal most significantly when the precession axis is not aligned along a high-symmetry direction of the sample. By correcting for these terms we achieve improved precision in measuring $d\tau/dV$, with consistent values at different $\theta$. We determine that the bias dependence of $d\tau_{\parallel}/dV$ is much weaker than reported by ref. [12] but still strong enough that it should be accounted for when analyzing experiments -- we find that $d\tau_\parallel/dV$ in a set of $RA = 12$ $\Omega$-$\mu m^2$ MTJs decreases by $35 \pm 10$ % between $V = -0.3$ V and 0.3 V, and $d\tau_\parallel/dV$ in a set of $RA$=1.5 $\Omega$-$\mu m^2$ MTJs decreases by $30 \pm 15$ % between $V = -0.15$ V and 0.15 V.

## EXPERIMENTAL METHODS AND ANALYSIS MODEL

In the ST-FMR technique,[6,7] a microwave-frequency current $I_{RF}$ and a direct current $I$ are applied to a MTJ. When the applied frequency, $f$, is close to a resonance of the magnetic layers, the magnetization in one or both of the electrodes can be driven to precess, giving rise to resistance oscillations. The experiment measures any DC voltage, $V_{mix}$, arising in response to $I_{RF}$. Near zero DC bias, the dominant resonant response is due to mixing between $I_{RF}$ and the resistance oscillations. The result is a resonant peak in the dependence of $V_{mix}$ on $f$; from the magnitude and peak shape for the lowest-frequency normal mode one can measure the in-plane and out-of-plane components of $d\boldsymbol{\tau}/dV$.[6,11] However, we note that in addition the applied $I_{RF}$ can cause the average (low-frequency) resistance of the MTJ to change. When $I \neq 0$, this effect should contribute an additional DC voltage signal, which must be taken into account when determining $d\boldsymbol{\tau}/dV$. Within a single-domain approximation, the leading-order contributions to $V_{mix}$ for small $I_{RF}$ are:

$$V_{mix} = \frac{1}{2}\frac{\partial^2 V}{\partial I^2}\left\langle\left(\delta I(t)\right)^2\right\rangle + \frac{\partial^2 V}{\partial I \partial \theta}\left\langle\delta I(t)\delta\theta(t)\right\rangle + \frac{1}{2}\frac{\partial^2 V}{\partial \theta^2}\left\langle\left(\delta\theta(t)\right)^2\right\rangle + \frac{\partial V}{\partial \theta}\left\langle\delta\theta(t)\right\rangle. \qquad (1)$$

Here $\delta I(t)$ and $\delta\theta(t)$ represent the full time dependence of the current and the offset angle between electrode magnetizations, relative to values for $I_{RF} = 0$, and $\langle\ \rangle$ denotes a time average. The first term in Eq. (1) is a non-resonant rectification background. The second term is the mixing voltage, the dominant resonant signal near zero DC bias. The third and fourth terms contribute only for non-zero bias, and describe, respectively, changes in the average low-frequency resistance due to the mean-square precession amplitude and due to a shift in the precession axis caused by $I_{RF}$. We do not take into account spin pumping,[8] because for MTJs this effect should be small compared to Eq. (1).

If we evaluate Eq. (1) assuming macrospin precession, with an initial magnetization orientation at any angle in the sample plane, we find to order $I_{RF}^2$ (see the Appendix):

$$V_{\text{mix}} = \frac{1}{4}\frac{\partial^2 V}{\partial I^2}I_{\text{RF}}^2 \tag{2a}$$

$$+\frac{1}{2}\frac{\partial^2 V}{\partial I\partial\theta}\frac{\hbar\gamma\sin\theta}{4eM_sVol\sigma}I_{\text{RF}}^2\big[\xi_\parallel S(\omega)-\xi_\perp\Omega_\perp A(\omega)\big] \tag{2b}$$

$$+\frac{1}{4}\frac{\partial^2 V}{\partial\theta^2}\left(\frac{\hbar\gamma\sin\theta}{4eM_sVol\sigma}\right)^2 I_{\text{RF}}^2\big(\xi_\parallel^2+\xi_\perp^2\Omega_\perp^2\big)S(\omega) \tag{2c}$$

$$+\frac{3}{8}\frac{\partial V}{\partial\theta}\frac{H_{\text{anis}}\sin2\beta}{(H_z+H_{\text{anis}}\cos2\beta)}\left(\frac{\hbar\gamma\sin\theta}{4eM_sVol\sigma}\right)^2 I_{\text{RF}}^2\big(\xi_\parallel^2+\xi_\perp^2\Omega_\perp^2\big)S(\omega) \tag{2d}$$

$$+\frac{\partial V}{\partial\theta}\frac{1}{M_sVol(H_z+H_{\text{anis}}\cos2\beta)}\left\{\frac{1}{4}\frac{\partial^2\tau_\perp}{\partial\theta^2}I_{\text{RF}}^2\right. \tag{2e}$$

$$+\frac{1}{2}\frac{\partial^2\tau_\perp}{\partial I\partial\theta}\frac{\hbar\gamma\sin\theta}{4eM_sVol\sigma}I_{\text{RF}}^2\big[\xi_\parallel S(\omega)-\xi_\perp\Omega_\perp A(\omega)\big] \tag{2f}$$

$$+\frac{1}{4}\frac{\partial^2\tau_\perp}{\partial\theta^2}\left(\frac{\hbar\gamma\sin\theta}{4eM_sVol\sigma}\right)^2 I_{\text{RF}}^2\big(\xi_\parallel^2+\xi_\perp^2\Omega_\perp^2\big)S(\omega)\left.\right\}. \tag{2g}$$

Here $\gamma$ ($> 0$) is the gyromagnetic ratio, $M_sVol$ is the total magnetic moment of the precessing layer, $\xi_\parallel = (2e/\hbar\sin\theta)(dV/dI)d\tau_\parallel/dV$ and $\xi_\perp = (2e/\hbar\sin\theta)(dV/dI)d\tau_\perp/dV$ represent the in-plane and out-of-plane torkances in dimensionless units, $H_{\text{anis}}$ is the within-plane anisotropy strength of the precessing layer, $H_z$ is the component of the magnetic field acting on the precessing layer along its equilibrium direction (including the applied external field and the dipole field but excluding the demagnetization field), and $\beta$ is the angle between the precessing layer's equilibrium direction and the magnetic easy axis [Fig. 1(a,c)]. $S(\omega)=[1+(\omega-\omega_m)^2/\sigma^2]^{-1}$ and $A(\omega)=[(\omega-\omega_m)/\sigma]S(\omega)$ are the symmetric and antisymmetric components of the lineshape, with $\omega_m$ the resonance frequency

$$\omega_m = \gamma M_{\text{eff}}\sqrt{N_x\left(N_y-\frac{1}{M_{\text{eff}}M_sVol}\frac{\partial\tau_\perp}{\partial\theta}\right)}, \tag{3}$$

$\sigma$ the linewidth

$$\sigma = \frac{\alpha\gamma M_{\mathrm{eff}}(N_x + N_y)}{2} - \frac{\gamma}{2M_s Vol}\frac{\partial\tau_\parallel}{\partial\theta}, \tag{4}$$

$N_x = 4\pi + (H_z - H_{\mathrm{anis}}\sin^2\beta)/M_{\mathrm{eff}}$ , $N_y = (H_z + H_{\mathrm{anis}}\cos 2\beta)/M_{\mathrm{eff}}$, and $\Omega_\perp = \gamma N_x M_{\mathrm{eff}}/\omega_m$, with $4\pi M_{\mathrm{eff}}$ the effective out-of-plane anisotropy for the precessing layer. Of the contributions in Eq. (2), only terms (2a) and (2b) were considered by Kubota et al. [12] In our previous experiment,[11] we discussed Eqs. (2a)-(2c), but we estimated that $\partial^2 V/\partial\theta^2$ was small for $\theta$ near 90° and therefore did not include Eq. (2c) in our final calculation of the torkances. Equation (2d) was zero for our previous geometry because $\beta$ was equal to 90°.[11]

## EXPERIMENTAL RESULTS

We have investigated devices from two different sets of magnetic tunnel junctions. We measured seven MTJs from the same batch of samples studied in ref. [11], with nominal $RA$ = 12 $\Omega$-$\mu m^2$ and the following layers (in nm) deposited onto an oxidized silicon wafer: bottom electrode [Ta(5)/Cu(20)/Ta(3)/Cu(20)], synthetic antiferromagnet (SAF) layer pinned to PtMn [PtMn(15)/Co$_{70}$Fe$_{30}$(2.5)/Ru(0.85)/Co$_{60}$Fe$_{20}$B$_{20}$(3)], tunnel barrier [MgO(1.25)] magnetic free layer [Co$_{60}$Fe$_{20}$B$_{20}$(2.5)], and capping layer [Ta(5)/Ru(7)]. The top (free) magnetic layer of these samples is etched to be a rounded rectangle, with dimensions either 50 × 100 nm$^2$ or 50 × 150 nm$^2$. The bottom SAF layers are left extended, with an exchange bias parallel to the magnetic easy axis of the top layer [Fig. 1(c)]. The insulator surrounding the sides of these devices is silicon oxide, and the top electrode is made using layers of Ta, Cu, and Pt. We also measured five MTJs with nominal $RA$ = 1.5 $\Omega$-$\mu m^2$ and the layer structure (in nm): bottom electrode [Ta(3)/CuN(41.8)/Ta(3)/CuN(41.8)/Ta(3)/Ru(3.1)], SAF layer pinned to IrMn [IrMn(6.1)/CoFe(1.8)/Ru/CoFeB(2.0)], tunnel barrier [MgO$_x$] free layer [CoFe(0.5)/CoFeB(3.4)], capping layer [Ru(6)/Ta(3)/Ru(4)]. In this second batch of samples, both the top magnetic free layer [the CoFe(0.5)/CoFeB(3.4) composite layer] and the bottom magnetic "pinned" layers [the CoFe(1.8)/Ru/CoFeB(2) SAF structure] are etched into a circular shape with diameter nominally 90 nm [Fig. 1(d)]. The insulator to the side of these devices is aluminum oxide, and top contact is made with Au. These samples are similar to the devices studied by Kubota et al.[12] both in the value of $RA$ and in that the pinned SAF electrode is etched. We will report data from a single $50 \times 100$ nm$^2$ sample (sample #1) of the first type with $RA$ = 12 $\Omega$-$\mu$m$^2$ and another sample (sample #2) of the second type with $RA$ = 1.5 $\Omega$-$\mu$m$^2$, but the results from all of the devices within each type were similar. Sample #1 had a zero-bias resistance of 3.9 k$\Omega$ in the parallel state and a tunneling magnetoresistance ratio of 160%. (This resistance is greater than for the sample in ref. [11], 3.19 k$\Omega$, with the consequence that the torkances we report for sample #1 are smaller than in ref. [11] by a factor of approximately 3.19/3.9 = 0.8.) Sample #2 had a zero-bias resistance of 279 $\Omega$ in the parallel state and a tunneling magnetoresistance ratio of 92%.

All of the ST-FMR measurements we report were performed at room temperature. Positive current is defined such that electrons flow from the top layer to the pinned layer. To generate different values of $\theta$, we apply an external magnetic field within the plane of the magnetic layers along various directions $\varphi$ (defined relative to the exchange-bias direction), selected to give well-separated resonances. We sweep the frequency $f$ of $I_{RF}$ while keeping the magnitude constant (< 10 $\mu$A for sample #1 and < 100 $\mu$A for sample #2), resulting in an average precession angle < 1°. The magnitude of $I_{RF}$ at the sample is calibrated using the non-resonant background [Eq. (2a)] with the procedure described in ref. [11]. The factors $\partial^2 V / \partial I^2$, $\partial^2 V / \partial I \partial \theta$, $\partial V / \partial \theta$, and $\partial^2 V / \partial \theta^2$ needed to calculate the torkance from $V_{mix}$ using Eq. (2) are determined for each sample by measuring $\partial V / \partial I$ over a range of biases and angles using a lock-in amplifier, integrating to determine $V$ vs. $I$, determining $\theta$ by assuming that the angular dependence of the zero-bias conductance is proportional to $\cos \theta$ and that $\theta$ does not change with bias, and then calculating the necessary terms numerically. (Within the voltage range we investigated, for $H \geq 250$ Oe, the DC spin torque due to the DC bias should change $\theta$ by less than 1° for sample #1 and less than 3° for sample #2.) We determine the anisotropy field $H_{anis}$ acting on the free layer and the exchange-bias field acting on the fixed layer by comparing measurements of $dV/dI$ vs. field angle $\varphi$ to macrospin simulations, and then use these simulations to determine the equilibrium direction $\beta$ of the free-layer moment. As in ref. [11], we use $M_sVol = 1.06 \times 10^{-14}$ emu ($\pm 15\%$), $M_S = 1100$ emu/cm$^3$, and $4\pi M_{eff} = 11 \pm 1$ kOe for sample #1. For sample #2 we use $M_sVol = 1.8 \times 10^{-14}$ emu ($\pm 15\%$) based on the measured value for the magnetization per unit area ($\langle M_S \rangle t = 3.2 \times 10^{-4}$ emu/cm$^2$) and our estimate of the sample area from scanning electron microscopy. The true area of the free layers in both types of devices is less than the nominal lithography dimensions because the sidewalls of the device are not vertical. We estimate $4\pi M_{eff} = 13 \pm 1$ kOe for sample #2 by comparing our measured FMR frequency to Eq. (3).

Figure 2(a) shows measured ST-FMR resonance peaks at selected values of $\theta$ for sample #1. In each spectrum we observe only a single large resonance. As predicted by Eq. (2), each resonance can be fit accurately by a sum of symmetric and antisymmetric Lorentzians ($S(\omega)$ and $A(\omega)$) with a frequency-independent background. For sample #2, in contrast to sample #1, we always observe two closely-spaced peaks in the ST-FMR spectra (Fig. 2(b)). We attribute this difference to the fact that the pinned electrode in sample #2 is etched, while the pinned magnetic electrode in sample #1 is left as an unetched extended film. This etching leaves the upper CoFeB layer within the CoFe/Ru/CoFeB SAF in sample #2 free to precess in response to a spin torque (in addition to the free layer), giving a second resonant mode. Coupling between the two modes has the potential to alter the magnitudes and the lineshapes of ST-FMR resonances in ways that are not included in our model. In an attempt to minimize such coupling effects, when analyzing the data from sample #2 we have selected values of magnetic field (both magnitude and angle) to maximize the frequency difference between the two resonances. However, we do not claim that coupling effects are entirely absent.

If we take into account only the direct mixing contribution [Eq. 2(b)] to the ST-FMR resonance (as was done in ref. [11] and ref. [12]), "uncorrected" in-plane and out-of-plane torkances $d\tau_\parallel/dV$ and $d\tau_\perp/dV$ can be determined separately from the frequency-symmetric and antisymmetric components of each resonance (Fig. 3). For sample #1, as the offset angle between the electrode magnetizations is varied from 58° to 131°, the in-plane component of this uncorrected torkance changes continuously from the form we reported previously[11] (approximately independent of bias for $|V| < 0.3$ V and increasing at higher bias for both signs of $V$) to a form that is strongly asymmetric in bias (increases sharply at negative bias), similar to the results of ref. [12]. The "uncorrected" torkances for sample #2, with the much lower value of $RA$, show a very similar evolution as a function of $\theta$. We therefore conclude that the dramatic differences between the two previous experimental results (references [11] and [12]) are a consequence of the use of different initial offset angles (50°-90° in our previous work,[11] 137° in ref. [12]). Moreover, we will argue below that the apparent variation as a function of offset angle shown in Fig. 3 does not reflect the true, corrected values of the spin-transfer torkances, but that it is an artifact of neglecting terms in Eq. (2) that become significant at large bias.

In Fig. 4 we plot estimates of the contributions to the ST-FMR signal of the terms in Eqs. (2c) and (2d) (both associated with changes in the DC resistance in response to $I_{RF}$) for sample #1, normalized by the part of the direct mixing contribution (Eq. (2b)) proportional to $S(\omega)$. The terms in Eqs. (2c) and (2d) are negligible for $|V| < 0.15$ V for sample #1, give ~10% corrections for $0.15$ V $< |V| < 0.3$ V, and can grow to be larger than the direct mixing contribution for $|V| > 0.4$ V. Both terms also depend strongly on the offset angle, with particularly large corrections for large $\theta$, near antiparallel alignment. The other three corrections in Eq. (2) [terms (2e)-(2g)] are generally negligible when $H \geq$ 1 kOe, but they may be as large as 20% of $V_{mix}$ under very weak fields and high bias. We find that Eqs. (2c) and (2d) have the correct bias dependence (both terms are asymmetric in bias) and sufficient magnitude to fully explain the strongly-asymmetric bias dependence seen for larger $\theta$ in Fig. 3(a). For the circular sample #2, Eq. (2d) is zero since $H_{anis}$ is negligible, but Eq. (2c) has a significant amplitude relative to the direct mixing contribution for $|V| > 0.1$ V, and can explain the large asymmetric dependence seen for large $\theta$ in Fig. 3(b).

An improved measurement of $d\boldsymbol{\tau}/dV$ as a function of bias can be obtained by including all of the terms in Eq. (2) in the analysis. After doing so, our revised measurements of the spin-transfer torkances are plotted in Fig. 5(a,b) for sample #1 and Fig. 5(c,d) for sample #2. Theory predicts that both $d\tau_\parallel/dV$ and $d\tau_\perp/dV$ should be proportional to $\sin\theta$, so when these quantities are normalized by $\sin\theta$ as in Fig. 5, they should collapse onto single curves for each sample.[13,15,16] We find that including all of the terms from Eq. (2) does improve the quality of the collapse for $(d\tau_\parallel/dV)/\sin\theta$ for a significant range of $|V|$. For sample #1 in the range $|V| < 0.3$ V the spread in values becomes less than $\pm 15\%$, comparable to the estimated uncertainty (see the inset in Fig. 5(a)). In Fig. 5(e) we show in more detail the degree to which the extracted values of $(d\tau_\parallel/dV)/\sin\theta$ in the range $|V| < 0.3$ V are modified for sample #1 when the contributions of the correction terms are accounted for. For sample #2, the quality of the data collapse is likewise significantly improved in the range $|V| < 0.15$ V.

For larger biases, for $|V| > 0.3$ V for sample #1 or for $|V| > 0.15$ V for sample #2, the corrected values of $(d\tau_\parallel/dV)/\sin\theta$ differ strongly from Fig. 3, but the results for different values of $\theta$ are not consistent. Moreover, at high bias for some values of $\theta$ there is no real-valued solution for $d\tau_\parallel/dV$ based on Eq. (2). We conclude from these results that the ST-FMR technique does not give reliable values of the torkances at very large biases. As can be seen in Fig. 4, at large $|V|$ the artifacts that result from the changing DC resistance [Eqs. (2c) and (2d)] grow rapidly to become larger than the mixing term [Eq. (2b)] from which the torkances are extracted. Therefore at high bias even small uncertainties in the calibrations of $\partial^2 V/\partial\theta^2$ (~10-20%) and $\partial^2 V/\partial I\partial\theta$ (5-20%) can prevent an accurate subtraction of the artifacts, and the desired mixing signal cannot be isolated. Effects of heating and inelastic scattering, which are not included in Eq. (2), might also affect the measurements for large $|V|$.

The primary discrepancy between the results of the previous ST-FMR experiments concerned the bias dependence of the in-plane torkance, $d\tau_\parallel/dV$. After our correction, we observe a weak bias dependence consistent for all angles in the bias ranges where our calibrations are accurate, with $(d\tau_\parallel/dV)/\sin\theta$ decreasing by $35 \pm 10$ % from $V = -0.3$ V to $V = 0.3$ V for sample #1, and with the same quantity decreasing by $30 \pm 15$ % from $V = -0.15$ V to $V = 0.15$ V for sample #2. This is a much weaker variation than for the "uncorrected" torkances at large values of $\theta$ (Figs. 3(a) and 3(c)), although it is slightly stronger than we reported in ref. [11].

These experimental results can be compared to several theoretical models. Theodonis and collaborators[13] have calculated the bias dependence of the spin transfer torques in MTJs within a tight-binding model of the electron bands, and Xiao, Bauer, and Brataas[14] have have calculated the torques within the Stoner model by scattering theory. In comparing to the experiments, these groups have focused to a significant extent on explaining the strongly asymmetric-in-bias dependence of the type present in the "uncorrected" curves for large $\theta$ in Fig. 5(a) and 5(c) and reported by Kubota et al.[12] As we have explained above, we argue that these strong asymmetries in the torkance are an artifact of neglecting significant terms in the analysis for the ST-FMR signal at high bias, and that the true values of the in-plane torkances are only weakly bias dependent in both types of MTJs that we have measured throughout the bias range in which the measurements are trustworthy. We do not claim that either the tight-binding or Stoner calculations are necessarily inaccurate, but we suggest that the parameter regimes in which they predict a strongly asymmetric bias dependence for $d\tau_\parallel/dV$ are not the correct regimes for analyzing the existing experiments. Heiliger and Stiles have calculated the bias dependence of the spin transfer torkances by an *ab initio* Green's function approach for an Fe/MgO/Fe MTJ.[15] They plotted $\tau_\parallel(V)$ and $\tau_\perp(V)$ in Fig. 4(a) of ref. [15]; we show the corresponding values of $d\tau_\parallel(V)/dV$ and $d\tau_\perp(V)/dV$ in Fig. 6 after converting to the same units we use for our experimental data and assuming the same device area as for sample #1 ($3.9 \times 10^3$ nm$^2$). The agreement between the form of the calculated bias dependence and the measurements is excellent, including even the existence of a small negative slope in the dependence of $d\tau_\parallel/dV$ on $V$. In the calculation, $d\tau_\parallel/dV$ decreases by $\sim$ 60% between -0.5 V and 0.5 V, the same relative slope per unit voltage measured for our sample #1 (a decrease of 35 ± 10 % between $V$ = -0.3 V and 0.3 V).

In regard to the absolute magnitude of the in-plane torkance, the average value that we measure for $(d\tau_\parallel/dV)/\sin\theta$ near $V = 0$ is 0.10 ± 0.02 $(\hbar/2e)$k$\Omega^{-1}$ for sample #1 and $1.1 \pm 0.2$ $(\hbar/2e)$k$\Omega^{-1}$ for sample #2. For a symmetric magnetic tunnel junction, the zero bias value of $(d\tau_\parallel/dV)/\sin\theta$ is predicted to be:[17]

$$\frac{d\tau_\parallel/dV}{\sin\theta} = \frac{\hbar}{4e}\frac{2P}{1+P^2}\left(\frac{dI}{dV}\right)_P ,\qquad(5)$$

where $(dI/dV)_P$ is the conductance for parallel magnetic electrodes. Evaluating Eq. (5) for a spin polarization factor $P = 67\%$ and $(dI/dV)_P = 3.9$ k$\Omega$ appropriate for sample #1 gives $(d\tau_\parallel/dV)/\sin\theta = 0.12$ $(\hbar/2e)$k$\Omega^{-1}$. Therefore for this sample our measured torkance at $V = 0$ agrees with Eq. (5) within the experimental uncertainty associated with our estimate of the sample volumes. The result of Heiliger and Stiles (Fig. 6) is also in good accord with Eq. (5): given the calculated polarization ($P \sim 1$) and $RA$ product ($\sim$14.5 $\Omega$-$\mu$m$^2$) of their junction, together with the device area of sample #1, we have $(dI/dV)_P = 0.27$ k$\Omega^{-1}$ so that Eq. (5) predicts $(d\tau_\parallel/dV)/\sin\theta = 0.13$ $(\hbar/2e)$k$\Omega^{-1}$, to be compared to the value of $0.14$ $(\hbar/2e)$k$\Omega^{-1}$ from the *ab initio* calculation. However, for our sample #2, using the values $P = 56\%$ and $(dI/dV)_P = 279$ $\Omega$, Eq. (5) predicts $(d\tau_\parallel/dV)/\sin\theta = 1.52$ $(\hbar/2e)$k$\Omega^{-1}$ at V = 0. This is approximately 40% larger than the value of the in-plane torkance extracted from the ST-FMR measurement for sample #2. While this difference could be interpreted as casting doubt on the prediction of Eq. (5) for the lower-*RA* tunnel junction devices, we suspect that the discrepancy is due to coupling between the free layer and the top CoFeB layer within the CoFe/Ru/CoFeB SAF "pinned layer" in sample #2. If we assume that the larger, lower-frequency resonance peaks in Fig. 2(b) that we use in analyzing the torkance correspond to the acoustic mode in which these two layers precess with the same phase, coupled motion of these layers would reduce the mixing voltage because the relative excitation angle would be reduced, thereby decreasing the size of the resistance oscillation. Coupling between the precessing layers may also be the reason that the measurements of $d\tau_\parallel(V)/dV$ at different values of $\theta$ for sample #2 in Fig. 5(c) show more of a spread than the corresponding data for sample #1 in Fig. 5(a). The degree of coupling via the magnetic dipole interaction should vary as a function of $\theta$.

In contrast to the in-plane component of the spin-transfer torkance, the bias dependence that we measure for the perpendicular component, $d\tau_\perp / dV$, displays only a negligible correction after including the additional terms in Eq. (2). In agreement with our previous results[11] and with Kubota et al.,[12] we find that to a good approximation $d\tau_\perp / dV \propto V$, so that after integrating we have $\tau_\perp(V) \approx A_0 + A_1 V^2$, with $A_0$ and $A_1$ constants (differing for different samples). The bias-dependent part of this torque in our experiments is in the $+\vec{m} \times \hat{M}_{\text{fixed}}$ direction for both signs of bias, meaning that the "effective field" on the precessing moment due to the spin-transfer torque is oriented antiparallel to $\hat{M}_{\text{fixed}}$. The magnitude of $d\tau_\perp / dV$ can become comparable to $d\tau_\parallel / dV$ at high bias, so that this in-plane torque may certainly be significant for technological applications.

A different bias dependence for the perpendicular torkance has recently been suggested by Li et al., based on the switching statistics of MTJs at high bias.[18] They argue that $d\tau_\perp / dV \propto |V|$, meaning that the bias-dependent part of $\tau_\perp(V)$ would change sign upon reversing the bias. However, Li et al. also noted that their data could in principle be explained by an alternative mechanism -- by a bias-dependent reduction of the within-plane magnetic anisotropy strength ($H_K$ in ref [18]) much stronger than one would expect from simple Ohmic heating. Li et al. argued that this scenario was unlikely, but more recent measurements by Sun et al.[19] suggest that indeed the within-plane magnetic anisotropy in MTJs can be much more strongly bias dependent than is expected from Ohmic heating. Therefore, in our opinion, the experiments of Li et al. are more likely to be explained by very strong variations in magnetic anisotropy rather than by a spin-transfer torkance of the form $d\tau_\perp / dV \propto |V|$.

Up to this point of our analysis, we have focused on the magnitudes and the lineshapes of the ST-FMR resonances. The linewidths and center frequencies can also provide valuable information. The linewidth, $\sigma$, can be related to the magnetic damping via Eq. (4). In Fig. 7(a) we present for sample #1 the bias dependence of the effective damping defined as $\alpha_{\text{eff}} = 2\sigma / [\gamma M_{\text{eff}}(N_x + N_y)]$. From Eq. (4), our macrospin model

predicts

$$\alpha_{\text{eff}} = \alpha - \frac{1}{M_{\text{eff}}(N_x + N_y)M_s Vol} \frac{\partial \tau_{\parallel}(\theta, I)}{\partial \theta} , \qquad (6)$$

which reduces to the Gilbert damping $\alpha$ at $V = 0$ since the spin-torque is zero. We find a value for the Gilbert damping $\alpha = 0.010 \pm 0.002$ for sample #1, consistent for all angles, a value in agreement with previous studies. [20],[21] Theory predicts that $\tau_{\parallel}(\theta, V) \propto const(V)\sin\theta$, so that the bias dependence of the effective damping should be small near $\theta = \pi/2$, negative for smaller angles, and positive for larger angles. The bias dependence predicted by Eq. (6) is shown in Fig. 7(a) by the shaded regions, which depict the $\pm$ 15% uncertainty associated with our determination of the sample volume. These center lines of these regions correspond to the average values $d\tau_{\parallel}/dV$ determined above for sample #1 in the region -0.3 V < V < 0.3 V, and we have assumed a simple linear extrapolation to higher values of |V|. We find the slope of $\alpha_{eff}$ vs. V does indeed increase as a function of $\theta$, qualitatively as expected, and passes through zero near $\theta = \pi/2$.

The bias dependent changes of the center frequency of the ST-FMR resonances for sample #1 is shown in Fig. 7(b) for the different offset angles, along with the values predicted by Eq. (3). In computing the predicted values, we have assumed that $d\tau_{\perp}/dV$ is linear in V over the entire bias range and that $d\tau_{\perp}/dV \propto \sin\theta$, we have used the average value of the slope determined above for sample #1, and then we integrated to determine $d\tau_{\perp}/d\theta$. We find that the measured frequency variation is in most cases much stronger than the variation expected to result from the measured value of the perpendicular torkance by itself, and is of a different functional form (Eq. (3) predicts a symmetric bias dependence). Our interpretation of this result is that the influence of the perpendicular torkance on the precession frequency is so small that other bias-dependent effects are dominant. For example, Petit et al. have shown that heating effects can produce significant bias-dependent shifts in the precession frequency for aluminum-oxide-based MTJs.[22] We suspect that a combination of Ohmic and Peltier heating might explain the bias dependence of the frequency seen in Fig. 7(b). Spin-transfer associated with lateral spin diffusion, which can increase the degree of spatial nonuniformity of the precessional mode for one sign of bias and decrease the spatial nonuniformity for the other sign could also provide an asymmetric contribution to the bias dependence to the precession frequency.[23,24,25] If we are correct that these other effects produce larger changes in the precession frequency than the perpendicular spin-transfer torkance, this would explain why previous experiments which attempted to determine the perpendicular spin-torque by measuring the bias dependence of the frequency have reached conclusions that conflict with the ST-FMR measurements.[22,26]

Readers may be concerned that heating or these other effects might also affect the ST-FMR magnitude or lineshape significantly, thereby making our analysis (Eq. (2)) inaccurate and invalidating our determination of the torkances. However, we have calculated that at low bias simple heating should produce a much weaker relative effect on the ST-FMR magnitude and lineshape than on the frequency, and that any heating-induced ST-FMR signal should produce a different bias dependence than is observed at low $|V|$ (see the Supplementary Material in ref. [11]). At higher biases, where Eq. (2) ceases to describe the experiments accurately, heating effects may well influence the ST-FMR magnitudes and lineshapes.

For completeness, figures 8(a) and 8(b) show the bias dependence of the effective damping and the center frequency for sample #2. The effective damping near zero bias is $\alpha = 0.014 \pm 0.002$, on the high end of the typical Gilbert damping 0.006-0.013 measured in CoFeB films.[20,21] Moreover, the measured bias dependence of the effective damping is stronger than what is predicted, based on the average values of the in-plane torkance measured for sample #2 and plotted in Fig. 5(c). We suggest that these differences result from the same cause that we invoked to explain why the measured in-plane torkance for sample #2 is less than the value predicted by Eq. (5): coupling between the free layer and the top layer of the SAF. The decreased value of the measured torkance can account for the most of the difference in slopes in Fig. 8(a), while the increased effective volume of the two coupled layers may contribute to an increased damping at zero bias. As for the bias dependence of the precession frequency [Fig. 8(b)], just as for sample #1 [Fig. 7(b)], the measured bias dependence is stronger than the changes expected due to the perpendicular component of the spin torque by itself.

## CONCLUSIONS

We have achieved improved measurements of the spin-transfer torkance, $d\tau/dV$, in MgO-based MTJs by using ST-FMR studies as a function of the offset angle $\theta$ and by accounting for all contributions to the ST-FMR signal of order $I_{RF}^2$. We show that discrepancies between two previous measurements resulted from using different $\theta$ and neglecting angle-dependent contributions to the ST-FMR signal caused by changes in the DC resistance in response to $I_{RF}$. We believe that a very strongly asymmetric bias dependence for $d\tau_\parallel/dV$ reported in ref. [12], which was claimed [12] to support the predictions of tight-binding calculations,[13] is an artifact of neglecting these contributions. After correcting for the additional terms, we find that the bias dependence of $d\tau_\parallel/dV$ for both sets of MTJs that we have measured is weaker but still strong enough to be technologically relevant, varying by 30-35% in the range $|V| < 0.3$ V for samples with $RA$ = 12 $\Omega$-$\mu m^2$ and $|V| < 0.15$ V for samples with $RA$ = 1.5 $\Omega$-$\mu m^2$. These results appear to be in good accord with the *ab initio* calculations of ref. [15]. For larger values of $|V|$, the artifacts in the ST-FMR signal become so dominant that the extraction of torkance values by this technique becomes unreliable. The out-of-plane component $d\tau_\perp/dV$ is less affected by the correction terms than $d\tau_\parallel/dV$. At high bias $d\tau_\perp/dV$ can become comparable to $d\tau_\parallel/dV$, so that both components should be taken into account when modeling spin-torque dynamics in magnetic tunnel junctions.

## ACKNOWLEDGMENTS

We thank Y. Nagamine, D. D. Djayaprawira, N. Watanabe, and K. Tsunekawa of Canon ANELVA Corp. and D. Mauri of Hitachi Global Storage Technologies (now at Western Digital Corp.) for providing junction thin film stacks that we used to fabricate the tunnel junctions, and C. Heiliger and M. D. Stiles for allowing us to use their data in Fig. 6. Cornell acknowledges support from ARO, ONR, DARPA, NSF (DMR-0605742), and the NSF/NSEC program through the Cornell Center for Nanoscale Systems. We also acknowledge NSF support through use of the Cornell Nanofabrication Facility/NNIN and the Cornell Center for Materials Research facilities.

# APPENDIX

## Derivation of the ST-FMR signal

This derivation is a generalized version of the calculation given in the supplemental material of ref. [11]. Here, we allow arbitrary orientations within the sample plane for the initial value of the free-layer magnetization and for the applied magnetic field (in ref. [11] we considered only the case that both were aligned along the hard in-plane magnetic axis), and we take into account corrections to the ST-FMR signal due to terms of the form $\frac{\partial \mathcal{V}}{\partial \theta}\langle \delta\theta(t)\rangle$ that were neglected in ref. [11].

We define the coordinate system as in Fig. 1(a,c). The orientation of the precessing layer moment is $\hat{m}$ and the orientation of the pinned-layer moment is $\hat{M}_{\text{fixed}}$. The x-axis is perpendicular to the thin film sample plane in the direction $\hat{M}_{\text{fixed}} \times \hat{m}$, the z-axis is along the equilibrium direction of $\hat{m}$, and the y-axis is perpendicular to both x and z axes such that $\hat{M}_{\text{fixed}} \cdot \hat{y} > 0$. As in ref. [11], we assume that $\hat{m}$ undergoes small angle precession. Because of the large magnetic anisotropy of the thin film sample, we have $|m_x| << |m_y|$ during the precession and $m_z \approx 1 - \frac{1}{2}m_y^2$, where $m_x$, $m_y$, $m_z$ are the three components of unit vector $\hat{m}$. If we define the angle between the magnetizations of the two electrodes as $\theta(t) = \theta + \delta\theta(t)$, then the time-dependent changes are given by $\delta\theta(t) = -m_y$ for small-angle precession. In response to the RF current $I(t) = I + \delta I(t)$ (where $\delta I(t) = I_{RF}\,\mathrm{Re}(e^{i\omega t})$), we write the oscillation of the free-layer moment as:

$$m_x = m_{x0} + \mathrm{Re}\left(m_{x1}e^{i\omega t}\right) + \mathrm{Re}\left(m_{x2}e^{2i\omega t}\right) + ...$$
$$m_y = m_{y0} + \mathrm{Re}\left(m_{y1}e^{i\omega t}\right) + \mathrm{Re}\left(m_{y2}e^{2i\omega t}\right) + ... \tag{A1}$$

Here $m_{x0}$ and $m_{y0}$ are real numbers, and $m_{x1}$, $m_{y1}$, $m_{x2}$, $m_{y2}$… are complex. We expect the oscillation to be harmonic to the first order in $I_{RF}$, so all of the coefficients except $m_{x1}$ and $m_{y1}$ should be at least second order in $I_{RF}$.

The voltage *V(t)* across the sample depends on the instantaneous value of the current *I(t)* and the angle *θ(t)*. The DC voltage signal produced by rectification in ST-FMR can be calculated by Taylor-expanding *V(t)* to 2nd order in $I_{RF}$ and taking the time average over one precession period:

$$V_{\text{mix}} = \frac{\partial V}{\partial I}\left\langle \delta I(t) \right\rangle + \frac{\partial V}{\partial \theta}\left\langle \delta \theta(t) \right\rangle + \frac{1}{2}\frac{\partial^2 V}{\partial I^2}\left\langle \left(\delta I(t)\right)^2 \right\rangle + \frac{\partial^2 V}{\partial I \partial \theta}\left\langle \delta I(t)\delta\theta(t) \right\rangle + \frac{1}{2}\frac{\partial^2 V}{\partial \theta^2}\left\langle \left(\delta\theta(t)\right)^2 \right\rangle$$

$$= 0 - \frac{\partial V}{\partial \theta}m_{y0} + \frac{1}{4}\frac{\partial^2 V}{\partial I^2}I_{RF}^2 - \frac{1}{2}\frac{\partial^2 V}{\partial I \partial \theta}I_{RF}\,\text{Re}(m_{y1}) + \frac{1}{4}\frac{\partial^2 V}{\partial \theta^2}\left| m_{y1} \right|^2 . \qquad (A2)$$

Within the macrospin approximation, the dynamics of $\hat{m}$ for the precessing layer can be calculated from the Landau-Lifshitz-Gilbert (LLG) equation with the addition of spin-transfer-torque terms,

$$\frac{d\hat{m}}{dt} = -\gamma\hat{m}\times\vec{H}_{\text{eff}} + \alpha\hat{m}\times\frac{d\hat{m}}{dt} - \gamma\frac{\tau_\parallel}{M_s Vol}\hat{y} - \gamma\frac{\tau_\perp}{M_s Vol}\hat{x} . \qquad (A3)$$

Here $\gamma = 1.76\times10^{11}\ \text{T}^{-1}\text{s}^{-1}$ is the absolute value of the gyromagnetic ratio, $\alpha$ is the phenomenological Gilbert damping constant, $M_s Vol$ is the total magnetic moment of the precessing layer, and $\vec{H}_{\text{eff}} = \vec{H}_{\text{demag}} + \vec{H}_{\text{dip}} + \vec{H}_{\text{app}}$ is the total effective field acting on the precessing layer, including the demagnetizing field, the dipole field from the pinned layer, and the external applied field. We assume that the dipole field and the external field both have orientations within the sample plane. We use $H_y$ and $H_z$ to denote the y and z components of the sum of these two fields, *i.e.* $\vec{H}_{\text{dip}} + \vec{H}_{\text{app}} = H_y\hat{y} + H_z\hat{z}$ . The demagnetizing field consists of a large perpendicular-to-the-plane component $-m_x 4\pi M_{\text{eff}}\hat{x}$ favoring an easy-plane anisotropy plus a smaller component $-m_{\text{hard}}H_{\text{anis}}\hat{d}_{\text{hard}}$ favoring the easy axis within the sample plane. Here $m_{\text{hard}} = -m_y\cos\beta + m_z\sin\beta$ is the component of $\hat{m}$ along the hard in-plane magnetic direction $\hat{d}_{\text{hard}} = -\hat{y}\cos\beta + \hat{z}\sin\beta$ , where $\beta$ is the angle between the precession axis and the magnetic easy axis [Fig. 1(a)]. The total demagnetizing field depending on the instantaneous direction of the precessing moment therefore has the form

$$\vec{H}_{\text{demag}} = -m_x 4\pi M_{\text{eff}} \hat{x} - \left(m_y \cos\beta - m_z \sin\beta\right)\left(\cos\beta\right)H_{\text{anis}}\hat{y} + \left(m_y \cos\beta - m_z \sin\beta\right)\left(\sin\beta\right)H_{\text{anis}}\hat{z}$$

$$(\text{A4})$$

At equilibrium ($m_x = m_y = 0,\ m_z = 1$), the total effective field must be along $z$ axis, giving the constraint $H_y + H_{\text{anis}}\sin\beta\cos\beta = 0$.

We expand the components of the spin-transfer-torque to second order as a function of current and offset angle:

$$\tau_\parallel = \tau_\parallel^0 + \frac{\partial \tau_\parallel}{\partial I}\delta I + \frac{\partial \tau_\parallel}{\partial \theta}\delta\theta + \frac{1}{2}\frac{\partial^2 \tau_\parallel}{\partial I^2}\delta I^2 + \frac{\partial^2 \tau_\parallel}{\partial I \partial \theta}\delta I \delta\theta + \frac{1}{2}\frac{\partial^2 \tau_\parallel}{\partial \theta^2}\delta\theta^2, \qquad (\text{A5})$$

with the analogous expression for $\tau_\perp$.

Expanding the LLG Equation (Eq. (A3)) to 2$^{\text{nd}}$ order in $m_x$ and $m_y$,

$$\frac{dm_x}{dt} = -\gamma N_y M_{\text{eff}} m_y - \frac{3}{2}m_y^2 \gamma H_{\text{anis}}\sin\beta\cos\beta - \alpha \frac{dm_y}{dt} - \frac{\gamma}{M_s Vol}\left[\frac{\partial \tau_\perp}{\partial I}I_{RF}\,\text{Re}(e^{i\omega t}) - \frac{\partial \tau_\perp}{\partial \theta}m_y\right]$$

$$- \frac{\gamma}{M_s Vol}\left[\frac{\partial^2 \tau_\perp}{\partial I^2}I_{RF}^2(1 + \text{Re}(e^{2i\omega t})) + \frac{1}{2}\frac{\partial^2 \tau_\perp}{\partial \theta^2}m_y^2 - \frac{\partial^2 \tau_\perp}{\partial I \partial \theta}I_{RF}\,\text{Re}(e^{2i\omega t})m_y\right], \qquad (\text{A6a})$$

$$\frac{dm_y}{dt} = -\gamma N_x M_{\text{eff}} m_x - m_x m_y \gamma H_{\text{anis}}\sin\beta\cos\beta + \alpha \frac{dm_x}{dt} - \frac{\gamma}{M_s Vol}\left[\frac{\partial \tau_\parallel}{\partial I}I_{RF}\,\text{Re}(e^{i\omega t}) - \frac{\partial \tau_\parallel}{\partial \theta}m_y\right]$$

$$- \frac{\gamma}{M_s Vol}\left[\frac{1}{4}\frac{\partial^2 \tau_\parallel}{\partial I^2}I_{RF}^2(1 + \text{Re}(e^{2i\omega t})) + \frac{1}{2}\frac{\partial^2 \tau_\parallel}{\partial \theta^2}m_y^2 - \frac{\partial^2 \tau_\parallel}{\partial I \partial \theta}I_{RF}\,\text{Re}(e^{2i\omega t})m_y\right], \qquad (\text{A6b})$$

where $N_x = 4\pi + \dfrac{H_z}{M_{\text{eff}}} - \dfrac{H_{\text{anis}}\sin^2\beta}{M_{\text{eff}}}$ and $N_y = \dfrac{H_z}{M_{\text{eff}}} + \dfrac{H_{\text{anis}}\cos 2\beta}{M_{\text{eff}}}$ for our sample geometry. Then, substituting Eq. (A1) into Eq. (A6), collecting the terms for different frequency components, and solving these equations for $m_{y1}$ and $m_{y0}$, we have

$$m_{y1} = \frac{\gamma I_{RF}}{2 M_s Vol}\frac{1}{(\omega - \omega_m - i\sigma)}\left[i\frac{\partial \tau_\parallel}{\partial I} + \frac{\gamma N_x M_{\text{eff}}}{\omega_m}\frac{\partial \tau_\perp}{\partial I}\right] \qquad (\text{A7a})$$

$$m_{y0} = -\frac{1}{N_y M_{\text{eff}}}\left[\frac{1}{M_s Vol}\left(\frac{1}{4}\frac{\partial^2 \tau_\perp}{\partial I^2}I_{RF}^2 - \frac{1}{2}\frac{\partial^2 \tau_\perp}{\partial I \partial \theta}\text{Re}(m_{y1})I_{RF} + \frac{1}{4}\frac{\partial^2 \tau_\perp}{\partial I^2}\left|m_{y1}\right|^2\right.\right.$$

$$+ \frac{3}{4} H_{\text{anis}} \sin \beta \cos \beta \left| m_{y1} \right|^2 \Bigg]. \tag{A7b}$$

Here, the resonance precession frequency is

$$\omega_m = \gamma M_{\text{eff}} \sqrt{N_x \left( N_y - \frac{1}{M_{\text{eff}} M_s Vol} \frac{\partial \tau_\perp}{\partial \theta} \right)} \approx \gamma M_{\text{eff}} \sqrt{N_x N_y} \tag{A8}$$

and the linewidth is

$$\sigma = \frac{\alpha \gamma M_{\text{eff}} (N_x + N_y)}{2} - \frac{\gamma}{2 M_s Vol} \frac{\partial \tau_\parallel}{\partial \theta}. \tag{A9}$$

After substituting Eq. (A7) into Eq. (A2), we reach the formula for the ST-FMR signal, Eq. (2) in the main text.

**REFERENCES**

1.  W. H. Butler, X.-G. Zhang, T. C. Schulthess, and J. M. MacLaren, Phys. Rev. B **63**, 054416 (2001).

2. J. Mathon and A. Umerski, Phys. Rev. B **63**, 220403(R) (2001).

3. S. S. P. Parkin et al., Nat. Mater. **3**, 862 (2004).

4.  S. Yuasa, T. Nagahama, A. Fukushima, Y. Suzuki, and K. Ando, Nat. Phys. **3**, 868 (2004).

5. J. Hayakawa et al., Jpn. J. Appl. Phys. **44**, L1267 (2005).

6. A. A. Tulapurkar et al., Nature **438**, 339 (2005).

7. J. C. Sankey et al., Phys. Rev. Lett. **96**, 227601 (2006).

8. J. N. Kupferschmidt, S. Adam, and P. W. Brouwer, Phys. Rev. B **74**, 134416 (2006).

9. A. A. Kovalev, G. E. W. Bauer, and A. Brataas, Phys. Rev. B **75**, 014430 (2007).

10. W. Chen, J.-M. L. Beaujour, G. de Loubens, A. D. Kent, and J. Z. Sun, Appl. Phys. Lett. **92**, 012507 (2008).

11. J. C. Sankey et al., Nature Physics **4**, 67 (2008), including Supplemental Material.

12. H. Kubota et a., Nature Physics **4**, 37 (2008), including Supplemental Material.

13. I. Theodonis, N. Kioussis, A. Kalitsov, M. Chshiev, and W. H. Butler, Phys. Rev. Lett. **97**, 237205 (2006); M. Chshiev, I. Theodonis, A. Kalitsov, N. Kioussis, W. H. Butler, IEEE Trans. Magn. **44**, 2543 (2008).

14. J. Xiao, G. E. W. Bauer and A. Brataas, Phys. Rev. B **77**, 224419 (2008).

15. C. Heiliger and M. D. Stiles, Phys. Rev. Lett. **100**, 186805 (2008).  Heiliger and Stiles have communicated to us that the convention for the sign of bias in Fig. 4(a) of their paper is the same as our convention.

16. J. C. Slonczewski, Phys. Rev. B **71**, 024411 (2005).

17. J. C. Slonczewski and J. Z. Sun, J. Magn. Magn. Mater. **310**, 169 (2007).

18. Z. Li, S. Zhang, Z. Diao, Y. Ding, X. Tang, D. M. Apalkov, Z. Yang, K. Kawabata, and Y. Huai, Phys. Rev. Lett. **100**, 246602 (2008).

19. J. Z. Sun, M. C. Gaidis, G. Hu, E. J. O'Sullivan, S. L. Brown, J. J. Nowak, P. L. Trouilloud, and D. C. Worledge, J. Appl. Phys. **105**, 07D109 (2009).

20. C. Bilzer, T. Devolder, J.-V. Kim, G. Counil, C. Chappert, S. Cardoso, and P. P. Freitas, J. Appl. Phys. **100**, 053903 (2006).

21. G. D. Fuchs, J. C. Sankey, V. S. Pribiag, L. Qian, P. M. Braganca, A. G. F. Garcia, E. M. Ryan, Zhi-Pan Li, O. Ozatay, D. C. Ralph, and R. A. Buhrman, Appl. Phys. Lett. **91**, 062507 (2007).

22. S. Petit, C. Baraduc, C. Thirion, U. Ebels, Y. Liu, M. Li, P. Wang, and B. Dieny, Phys. Rev. Lett. **98**, 077203 (2007).

23. K.-J. Lee, arXiv:0811.4649.

24. M. L. Polianski and P. W. Brouwer, Phys. Rev. Lett. **92**, 026602 (2004).

25. M. D. Stiles, J. Xiao, and A. Zangwill, Phys. Rev. B **69**, 054408 (2004).

26. A. M. Deac, A. Fukushima, H. Kubota, J. Maehara, Y. Suzuki, S. Yuasa, Y. Nagamine, K. Tsunekawa, D. D. Djayaprawira, and N. Watanabe, Nature Phys. **4**, 803 (2008).

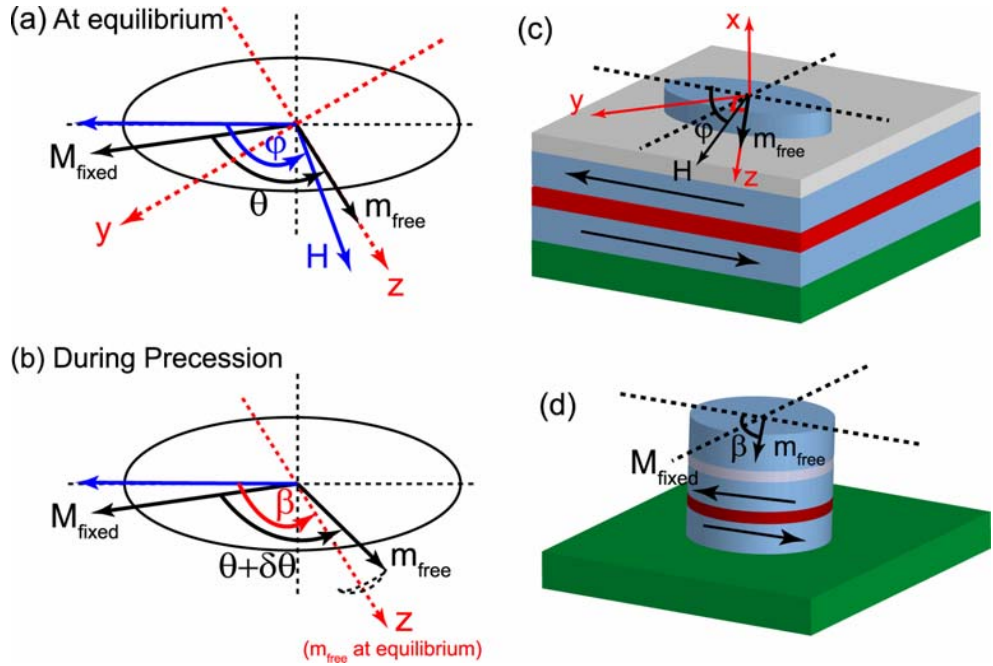

Fig. 1. (color online) (a) Definition of the coordinate system. The z-axis is defined as the equilibrium direction of the precessing-layer moment $m_{\text{free}}$. (b) Schematic of the free layer precession. The precession axis may be slightly misaligned from the z-axis when the last term in Eq. (1) is considered. (c) Schematic geometry for our samples with $RA =$ 12 $\Omega$-$\mu$m$^2$. The free layer is etched into a rounded rectangle while the bottom pinned layers are left extended. (d) Schematic geometry for our samples with $RA =$ 1.5 $\Omega$-$\mu$m$^2$. The synthetic antiferromagnetic pinned layers are etched, as well as the top free layer.

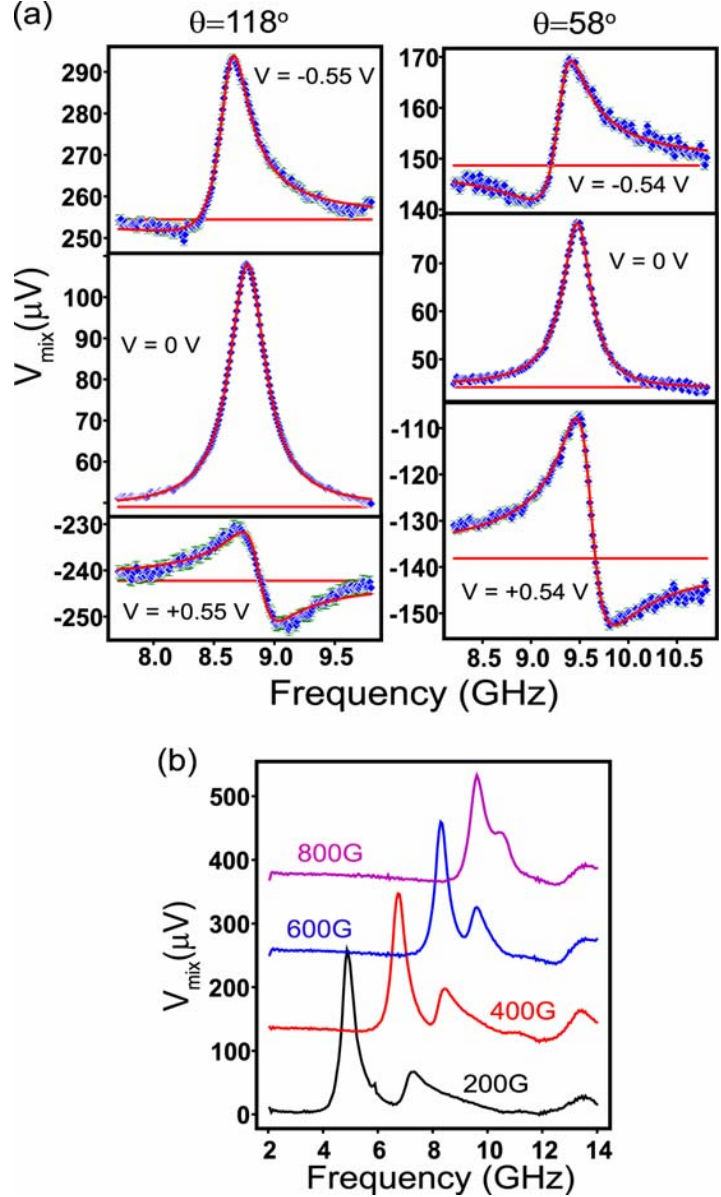

Fig. 2. (color online) (a) Measured ST-FMR spectra from sample #1 at negative, zero and positive biases for offset angles of θ = 115° and 58°, with fits to sums of symmetric and antisymmetric Lorentzians. (b) Measured ST-FMR spectra from sample #2 at zero bias with magnetic fields of various magnitudes applied in the φ = 130° direction. The spectra for sample #2 show two closely-spaced peaks, suggesting the existence of precessional dynamics in both the free magnetic layer and the etched synthetic antiferromagnet pinned layer.

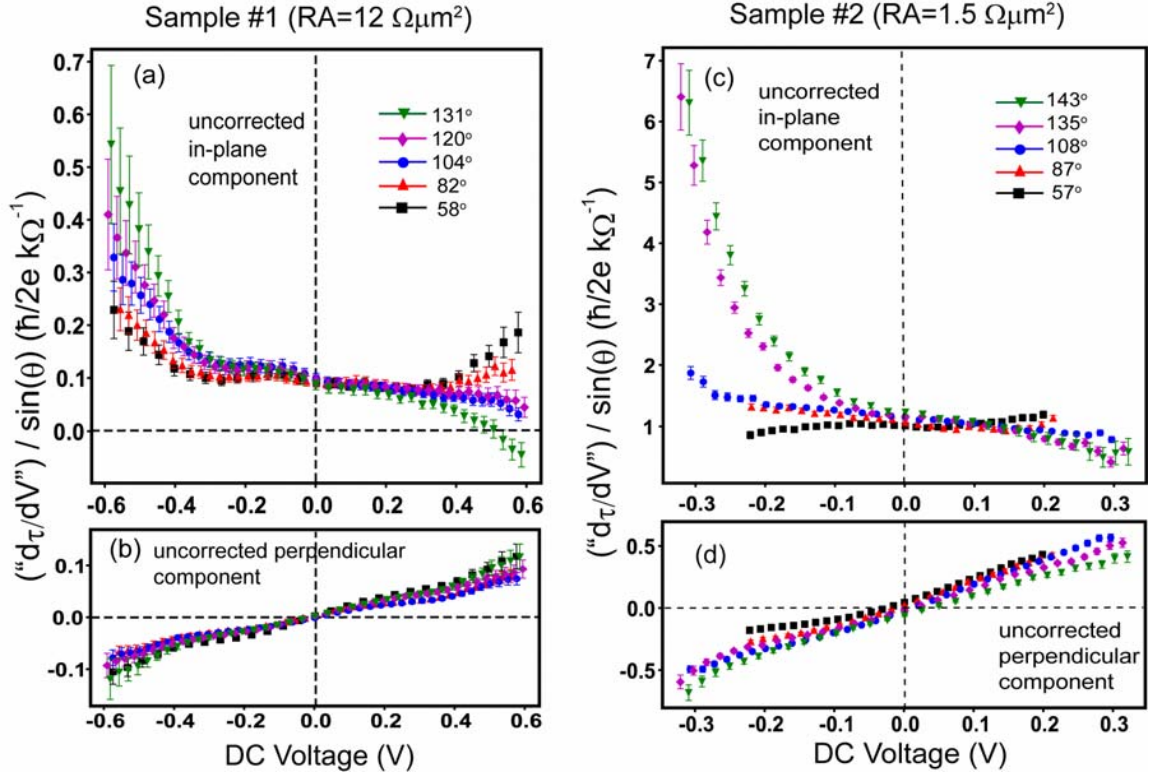

Fig. 3. (color online) Bias dependence of the uncorrected (a) in-plane and (b) out-of-plane torkances for sample #1 and uncorrected (c) in-plane and (d) out-of-plane torkances for sample #2 at several different offset angles $\theta$ respectively. (See the text.) The different offset angles are achieved by different combinations of applied field magnitude and direction. For sample #1, for $\theta = 131°$, $H = 0.38$ kOe with $\varphi = 120°$, giving $\beta = 142°$; for $\theta = 120°$, $H = 0.40$ kOe with $\varphi = 110°$, giving $\beta = 133°$; for $\theta = 104°$, $H = 0.75$ kOe with $\varphi = 130°$, giving $\beta = 142°$; for $\theta = 82°$, $H = 1.00$ kOe with $\varphi = 120°$, giving $\beta = 129°$; for $\theta = 58°$, $H = 1.00$ kOe with $\varphi = 90°$, giving $\beta = 96°$. For sample #2, for $\theta = 143°$, $H = 0.25$ kOe with $\varphi = 150°$; for $\theta = 135°$, $H = 0.25$ kOe with $\varphi = 140°$; for $\theta = 108°$, $H = 0.30$ kOe with $\varphi = 125°$; for $\theta = 87°$, $H = 0.20$ kOe with $\varphi = 85°$; for $\theta = 57°$, $H = 0.30$ kOe with $\varphi = 55°$. The value of $\beta$ is not needed in the calculations for sample #2 because its cross section is circular.

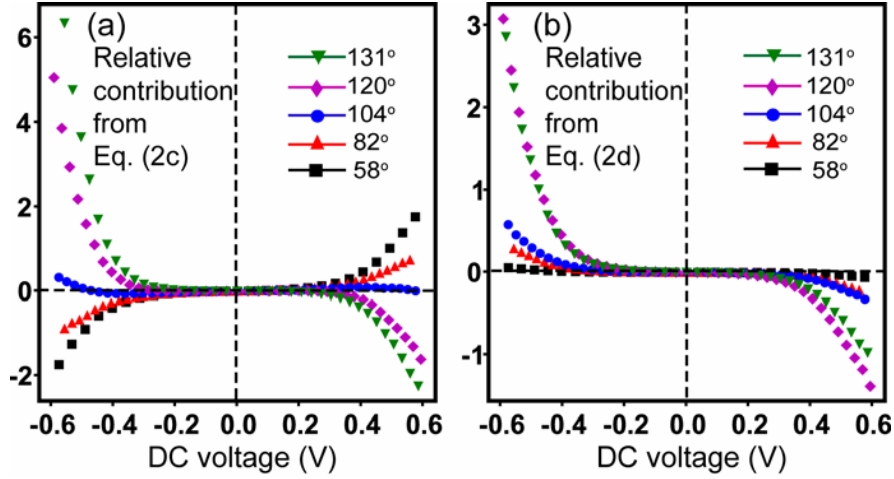

Fig. 4. (color online) Estimated contribution to the ST-FMR signal for sample #1 from (a) the term in Eq. (2c) and (b) the term in Eq. (2d), relative to the frequency-symmetric part of the direct mixing signal, Eq. (2b). The angles in the legends are the initial offset angles $\theta$. In both (a) and (b), we assume for simplicity that the in-plane torkance is a constant $d\tau_{\parallel}/dV = 0.10\ (\hbar/2e)k\Omega^{-1}$ and the perpendicular torkance has a constant slope $d^2\tau_{\perp}/dV^2 = 0.16\ (\hbar/2e)k\Omega^{-1}/V$.

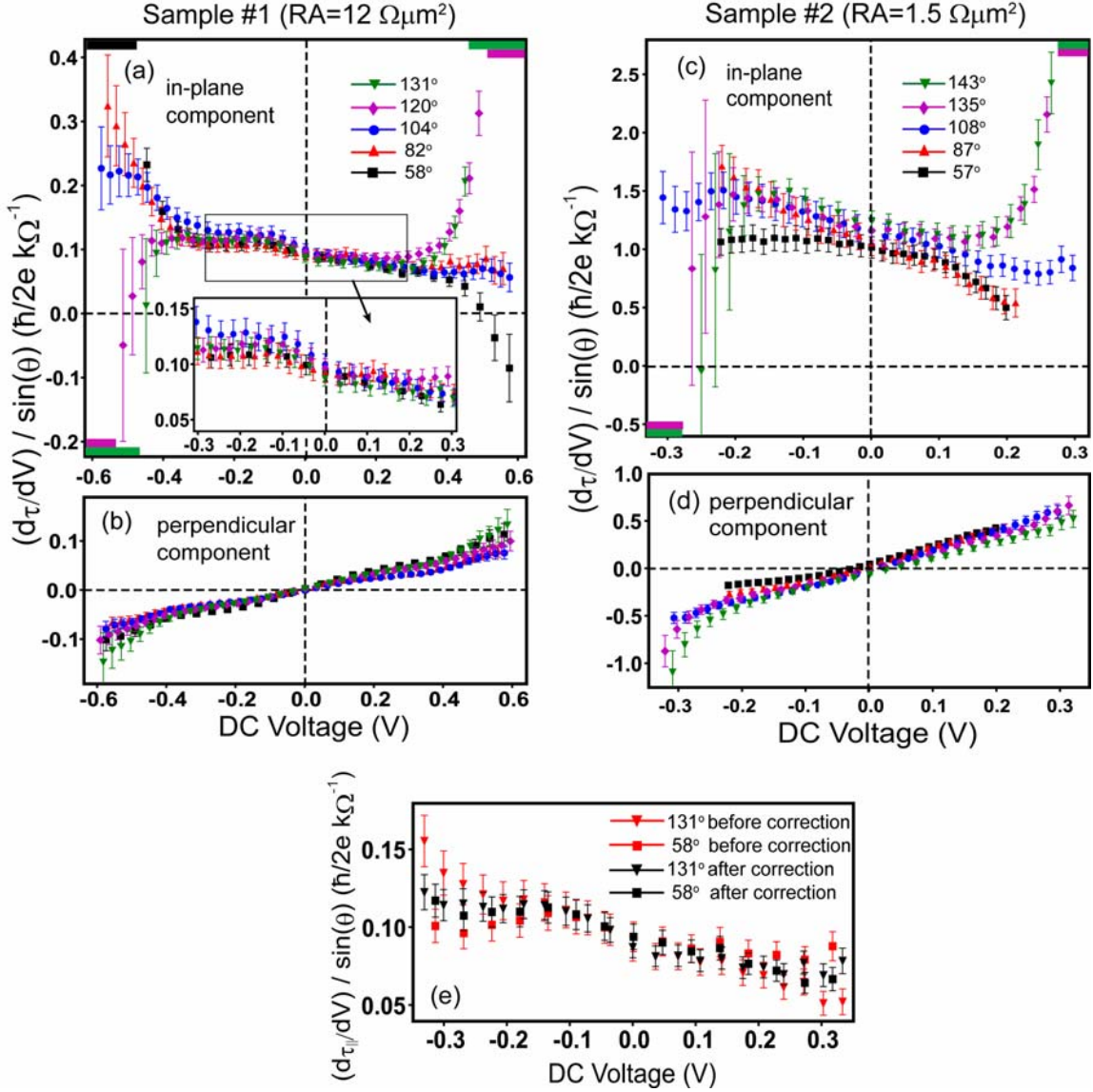

Fig. 5. (color online) Determinations of the (a) in-plane and (b) out-of-plane component of the spin-transfer torkance for sample #1 and the (c) in-plane and (d) out-of-plane component of the spin-transfer torkance for sample #2, from an analysis that includes all of the contributions to the ST-FMR signal in Eq. (2). The angles in the legend are the initial offset angles $\theta$. The corresponding values of $H$, $\varphi$ and $\beta$ are listed in the caption of Fig. 3. At the largest values of $|V|$ for some angles there is no real-valued solution for $d\tau_{\parallel}/dV$ based on Eq. (2) and our ST-FMR data, and we have marked these regimes with bars along the top or bottom axes in (a) and (c). (e) Uncorrected and corrected determinations of the in-plane component of the torkance for sample #1 for two values of the initial offset angle between the electrode magnetizations, $\theta = 58°$ and $131°$. Note that there is better consistency between the measurements for the two angles after the correction.

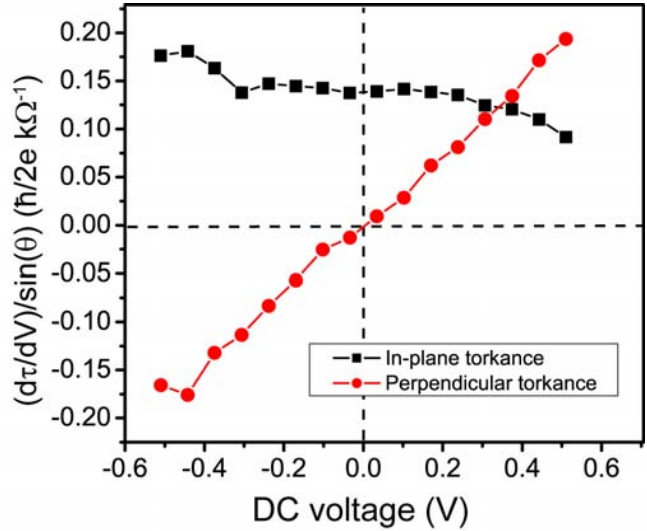

Fig. 6. In-plane and perpendicular torkances for an $RA \approx 14.5$ $\Omega$-$\mu m^2$ Fe/MgO/Fe tunnel junction calculated in ref. [15] using an *ab initio* multiple-scattering Green's function approach. The points are determined by numerical differentiation of the data in Fig. 4(a) of ref. [15]. We have converted to the units we use in describing the experiments assuming that the device area is the same as for sample #1 ($3.9 \times 10^3$ nm$^2$).

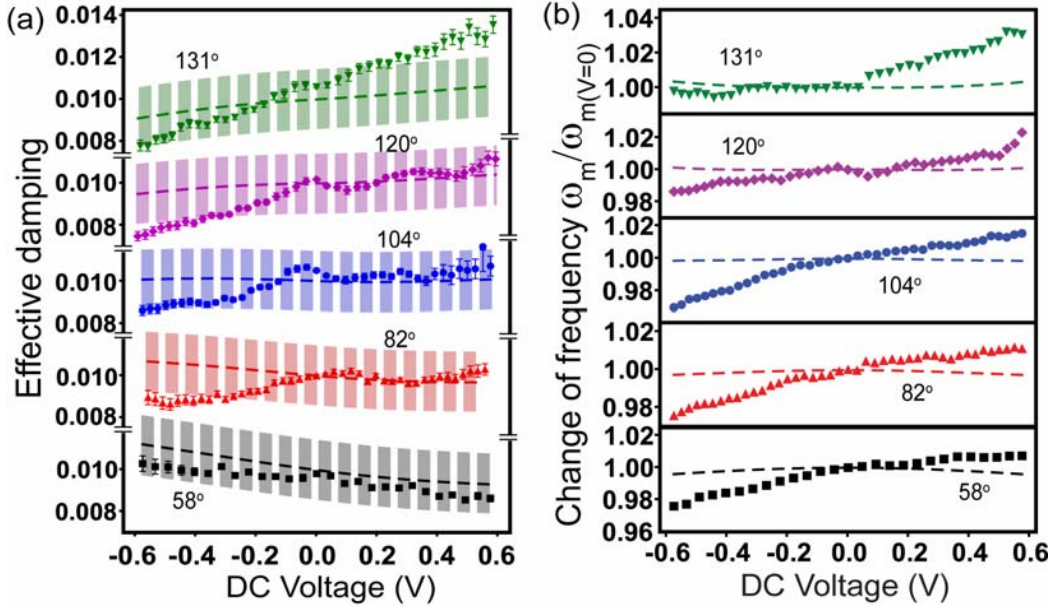

Fig. 7. (color online) (a) Bias dependence of the effective damping for various initial offset angles $\theta$ for sample #1. The corresponding values of $H$, $\varphi$ and $\beta$ are listed in the caption of Fig. 3. The dashed lines represent the effective damping predicted by Eq. (6), with the Gilbert damping constant $\alpha = 0.010$ and a fit to the measured in-plane spin-transfer torkance. The shaded regions reflect a ±15% uncertainty in the value of $M_s Vol$. (b) The bias dependent change of the resonance frequency $\omega_m$ for various initial offset angles $\theta$ for sample #1. The lines show the bias-dependent changes predicted by Eq. (3), using the measured value of the perpendicular component of the spin-transfer torkance. The measured bias dependence is much greater than the small variation expected from Eq. (3), suggesting that other effects (e.g., heating) may dominate the bias dependence of $\omega_m$, rather than spin torque being the only significant effect. The measured resonance frequencies at zero bias are: 9.47 GHz for $\theta = 58°$, 9.68 GHz for $\theta = 82°$, 8.93 GHz for $\theta = 104°$, 5.81 GHz for $\theta = 120°$, and 5.91 GHz for $\theta = 131°$.

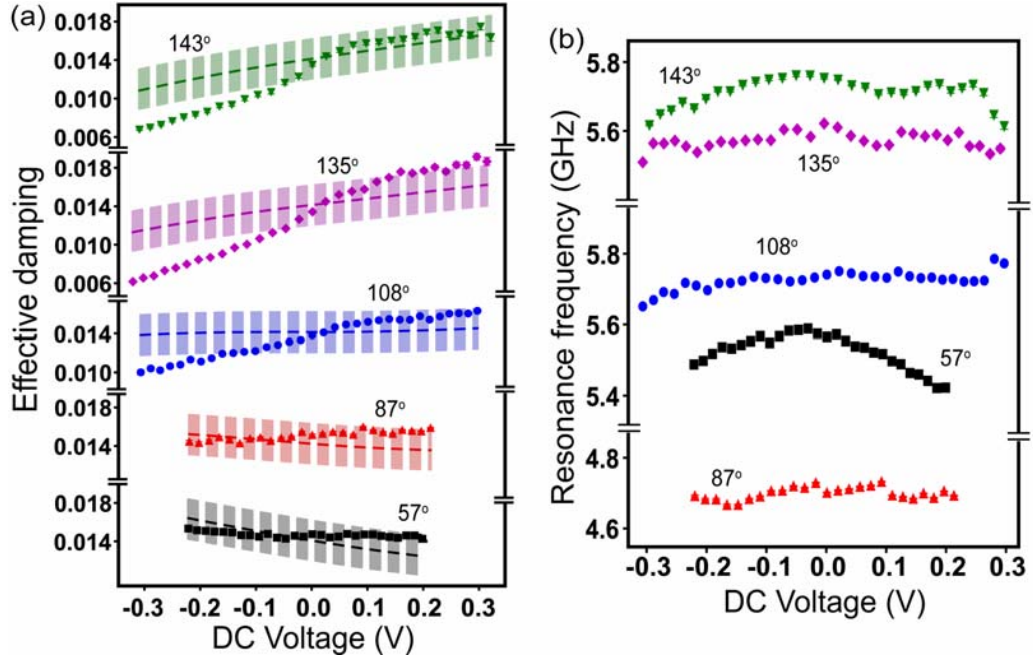

Fig. 8. (color online) (a) Bias dependence of the effective damping for various initial offset angles $\theta$ for sample #2. The dashed lines represent the effective damping predicted by Eq. (6), with the Gilbert damping constant $\alpha = 0.014$ and a fit to the measured in-plane spin-transfer torkance. The shaded regions reflect a ±15% uncertainty in the value of $M_s Vol$. (b) The bias dependence of the resonance frequency $\omega_m$ for various initial offset angles $\theta$ for sample #2.